\documentclass[twocolumn,prl,amsmath,amssymb,floatfix,superscriptaddress]{revtex4-2}

\usepackage{mathtools}
\usepackage[english]{babel}
\usepackage{amsfonts}
\usepackage{amsmath}
\usepackage{bm}
\usepackage{bbold}
\usepackage{tikz}
\usepackage{xcolor}
\usepackage[colorlinks, citecolor={blue}, urlcolor={blue}, linkcolor={red!80!black}]{hyperref}
\usepackage{orcidlink}

\usepackage{siunitx}
\usepackage{physics}
\ExplSyntaxOn
\msg_redirect_name:nnn{siunitx}{physics-pkg}{none} % remove warning because siunitx and physics both define the \qty command
\ExplSyntaxOff

\begin{document}

\title{Generalized conditions for odd-frequency pairing in superconducting systems}

\author{{Florian Kayatz}\,\orcidlink{0009-0005-3572-3561}}
\email{florian.kayatz@physics.uu.se}
\affiliation{Department of Physics and Astronomy, Uppsala University, Box 524, SE-751 20 Uppsala, Sweden}
\affiliation{Wallenberg Initiative Materials Science for Sustainability, Uppsala University, SE-751 20 Uppsala, Sweden}

\author{{Annica M. Black-Schaffer}\,\orcidlink{0000-0002-4726-5247}}
\affiliation{Department of Physics and Astronomy, Uppsala University, Box 524, SE-751 20 Uppsala, Sweden}
\affiliation{Wallenberg Initiative Materials Science for Sustainability, Uppsala University, SE-751 20 Uppsala, Sweden}

\author{{Jorge Cayao}\,\orcidlink{0000-0001-6037-6243}}
\affiliation{Department of Physics and Astronomy, Uppsala University, Box 524, SE-751 20 Uppsala, Sweden}

\date{\today}

%------------------------------------------------------------------------
% Abstract
%------------------------------------------------------------------------
\begin{abstract}
    Odd-frequency superconducting pairing has been predicted to arise in many systems and is known to lead to phenomena such as paramagnetic Meissner response and long-range superconducting proximity effect. Here, we provide generalized necessary and sufficient conditions for odd-frequency pairing appearing in any superconducting system, from bulk superconductors to superconducting hybrid structures. This generalizes an earlier first-order expression in multiband bulk superconductors to all superconducting systems and to all orders, in both order parameter and frequency.
    We then apply the derived conditions to several systems, including superconducting-ferromagnet and superconducting Josephson junctions, as well as a transition metal dichalcogenide monolayer proximitized by a conventional superconductor, where the  generalized conditions are used to understand the properties of the superconducting state.
\end{abstract}

\maketitle

\begin{figure}[!t]
    \centering
    \includegraphics[width=\columnwidth]{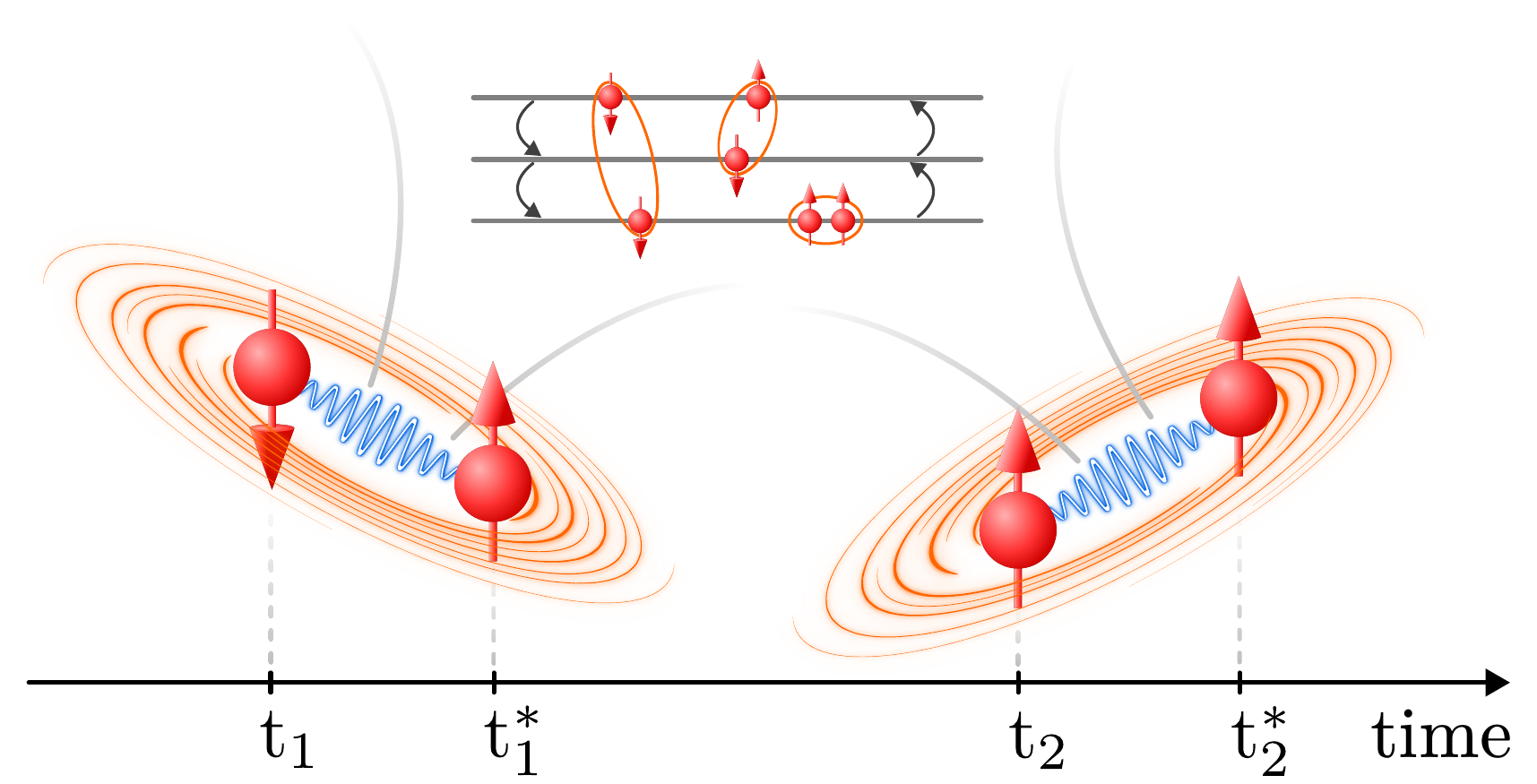}
    \caption{Sketch illustrating odd-frequency pairing, formed by an interaction at different times $t_i$ and $t_i^*$. Odd-frequency pairing states are encountered when additional degrees of freedom are present, here indicated by multiple bands.}
    \label{fig:oddfreqsketch}
\end{figure}

%------------------------------------------------------------------------
% SECTION I: Introduction
%------------------------------------------------------------------------
Odd-frequency Cooper pairs have come to play a key role in understanding superconducting phases \cite{RevModPhys.77.935,RevModPhys.77.1321,JPSJ.81.011013,Eschrig_2015,linder_superconducting_spintronics_2015,linder_odd-frequency_2019,Cayao2020,triola_role_2020},
with their exotic nature being shown to be directly connected to emergent phenomena, such as spin-triplet superconducting states \cite{PhysRevB.68.064513,PhysRevB.90.220501,PhysRevB.92.134512,PhysRevB.92.205424,PhysRevLett.116.257001,PhysRevB.98.075425,PhysRevLett.120.037701,PhysRevB.100.104511,Maeda2025,PhysRevB.111.064502,chakraborty2024,c325-kgbf,FukayaJPCM2025,lkf9-jgv6,SciPostPhys.20.2.059}, Majorana zero modes \cite{PhysRevB.72.140503,PhysRevB.87.104513,PhysRevB.92.100507,PhysRevB.95.184506, PhysRevB.95.174516,PhysRevB.96.155426,PhysRevB.101.094506,PhysRevB.101.214507, PhysRevB.101.195303,10.1093/ptep/ptae065,PhysRevB.110.125408,fksg-x8pr,PhysRevB.111.024507,ahmed2025anomalous,sardinero2026odd}, unusual Meissner response \cite{PhysRevB.52.1271,PhysRevB.89.184508, PhysRevX.5.041021, PhysRevB.95.184506, PhysRevMaterials.5.114801, c57s-skv9},
impurity responses \cite{PhysRevLett.125.117003,PhysRevB.103.024501,PhysRevLett.129.247001},
and long-range supercurrents \cite{petrashov1994conductivity,A.Kadigrobov_2001,PhysRevLett.86.4096,PhysRevB.68.064513,RevModPhys.77.935,RevModPhys.77.1321,Keizer2006,PhysRevLett.98.077003,PhysRevLett.99.127002,PhysRevLett.104.137002,science.1189246}, with the latter being a key to superconducting spintronics \cite{Eschrig_2015, linder_superconducting_spintronics_2015, yang_superconducting_spintronics_2021,mel2022superconducting}.
Specifically, odd-frequency Cooper pairs exhibit a pair amplitude that is an odd function in the relative time (or frequency) between the constituent electrons, as illustrated in Fig.~\ref{fig:oddfreqsketch}, a peculiar dynamical concept that was originally proposed by Berezinskii for superfluid $^3\mathrm{He}$ \cite{osti_4213537}. Such odd-frequency pair amplitudes are allowed to appear as soon as an overall full antisymmetry condition, derived from Fermi-Dirac statistics of the electrons forming Cooper pairs, is satisfied \cite{linder_odd-frequency_2019}.

It is by now also well-understood that odd-frequency Cooper pairs occur when breaking symmetries linked to the degrees of freedom of the paired electrons, for example due to magnetic fields \cite{RevModPhys.77.1321}, interfaces or junctions \cite{JPSJ.81.011013,Cayao2020}, time-periodic drives \cite{PhysRevB.103.104505,PhysRevB.109.134517,lkf9-jgv6,SciPostPhys.20.2.059}, or when multiple unequal bands/orbitals are present \cite{triola_role_2020,PhysRevB.88.104514,PhysRevB.92.054516,PhysRevB.92.094517, PhysRevB.92.094517,PhysRevB.93.201402,PhysRevLett.116.257001,PhysRevLett.119.087001,PhysRevB.97.214508,PhysRevB.109.205406,chakraborty2026breathing}.
In multiband bulk superconductors, the occurrence of odd-frequency pairing has even been formulated, to leading order in the order parameter $\hat{\Delta}$, as a kind of non-commuting relationship between $\hat{\Delta}$ and the normal state Hamiltonian $\hat{h}$ \cite{triola_role_2020}
\begin{equation}
    [\hat{h},\hat{\Delta}]_* \equiv \hat{h} \hat{\Delta} - \hat{\Delta} \hat{h}^* \neq 0\,. \label{eq:odd_freq_cond_triola}
\end{equation}
Notably, this condition is equivalent to the superconducting fitness in bulk superconductors \cite{PhysRevB.94.104501, ramires_fitness_2018, Fischer_2013}, which gives a suppression of the critical temperature. This highlights the importance of controlling odd-frequency pairing to both understand and engineer the superconducting state. Yet the universal applicability of Eq.~\eqref{eq:odd_freq_cond_triola} to other superconducting systems remains an open question.

In this work, we generalize the condition in Eq.~\eqref{eq:odd_freq_cond_triola} and show that it is both necessary and sufficient for odd-frequency pairing, at any order in the order parameter and, importantly, valid in {\it all} superconducting systems, from bulk to hybrid structures.
We further show that a high-frequency expansion provides computationally efficient access to the internal orbital structure of the odd-frequency Cooper pairs.
% We also show that an additional high-frequency expansion provides computationally efficient access to the internal orbital structure of the odd-frequency Cooper pairs.
We illustrate the utility of our results in three representative systems:
a superconductor-ferromagnet junction, where the conditions reveal the necessity of a spin-active interface for odd-frequency equal-spin pairing;
a Josephson junction under a magnetic field, where we show that the superconducting phase difference enables control over the formation of odd-frequency equal-spin pairing;
and a transition metal dichalcogenide monolayer proximitized by a conventional superconductor, where our generalized conditions help design the orbital nature of proximity-induced odd-frequency multiband pairing.
Our results thus offer a generalized framework for understanding and engineering odd-frequency Cooper pairs in any superconducting system.

%------------------------------------------------------------------------
% SECTION II: General condition for odd-frequency pairing
%------------------------------------------------------------------------
\textit{\label{sec:prev_cond}General framework.}---We start by describing any superconducting system with a Bogoliubov-de Gennes (BdG) Hamiltonian
\begin{equation}
    \mathcal{H} = \frac{1}{2} \Psi^\dagger H_{\text{BdG}} \Psi\,,\quad
    H_{\text{BdG}} = \begin{pmatrix}
        \hat{h}              & \hat{\Delta} \\
        \hat{\Delta}^\dagger & -\hat{h}^*
    \end{pmatrix}, \label{eq:ham_bdg}
\end{equation}
where $\hat{h}$ is the normal-state Hamiltonian and $\hat{\Delta}$ is the superconducting order parameter, assumed non-zero in at least a part of the system. Here, ${\Psi^\dagger = (\hat{\Phi}^\dagger, \hat{\Phi}^T)}$ is given in the basis $\hat{\Phi}^T = \{c_1,\ldots,c_n\}$ of the normal-state Hamiltonian, with $c_i$ annihilating an electron in the state $i$ and $i$ encoding all relevant degrees of freedom, including spin, orbital, and lattice indices.
Both $\hat{h}$ and $\hat{\Delta}$ are, in general, matrices, which also easily describe hybrid structures, where only part of the system host non-zero components of $\hat{\Delta}$.
When working in momentum space, the Hamiltonian is given by the substitutions ${\hat{h} \to \hat{h}_{\bm{k}}}$, ${\hat{h}^* \to \hat{h}^*_{-\bm{k}}}$ and ${\hat{\Delta} \to \hat{\Delta}_{\bm{k}}}$.
All conditions derived in this work remain valid under these replacements.

To characterize the superconducting correlations, we calculate the anomalous electron-hole Green's functions \cite{zagoskin,mahan2013many} for the BdG Hamiltonian in Eq.~\eqref{eq:ham_bdg} as
\begin{equation}
    F_{ij}(\tau) = -\langle \hat{T}_\tau c_{i}(\tau) c_{j}(0) \rangle\,, \label{eq:ftau}
\end{equation}
where $\tau$ is imaginary time and $\hat{T}_\tau$ the time-ordering operator.
In practice, we Fourier transform the total Green's function, calculated from the equation of motion as
\begin{equation}
    \label{eq:GF}
    \left(z - H_{\text{BdG}}\right) G(z)  = \mathbb{1}\,,\quad G(z)=\begin{pmatrix}
        G_0(z)     & F(z)         \\
        \bar{F}(z) & \bar{G}_0(z)
    \end{pmatrix},
\end{equation}
where $z$ represents complex frequencies, $G_0$ ($\bar{G}_0$) is the normal electron-electron (hole-hole) Green's function, and $F$ ($\bar{F}$) is the anomalous electron-hole (hole-electron) Green's function. The individual matrix elements of $F$ are then given by the Fourier transform of Eq.~\eqref{eq:ftau}.

Under the assumption that $ (z + \hat{h}^*) $, the Schur complement of $ (z + \hat{h}^*) $, and $(z - \hat{h})$ are all invertible, we write the anomalous part of the Green's function as
\begin{equation}
    \label{eq:f_generic_term}
    \begin{split}
        F(z) & = {\left[ 1 - {(z - \hat{h})}^{-1} \hat{\Delta} {(z + \hat{h}^*)}^{-1}
        \hat{\Delta}^\dagger \right]}^{-1}                                            \\
             & \quad \cdot {(z - \hat{h})}^{-1} \hat{\Delta} {(z + \hat{h}^*)}^{-1}.
    \end{split}
\end{equation}
Applying the standard assumption that superconductivity is the smallest energy scale, we then expand the first inverse in a Neumann series to obtain contributions at different orders of $\hat{\Delta}$:
\begin{equation}
    \label{eq:f_neumann_series}
    \begin{split}
        F(z) & =  \sum_{k=0}^{\infty} \left[ (z - \hat{h})^{-1} \hat{\Delta} (z + \hat{h}^*)^{-1} \hat{\Delta}^\dagger \right]^k \\
             & \quad \cdot (z - \hat{h})^{-1} \hat{\Delta} (z + \hat{h}^*)^{-1}.
    \end{split}
\end{equation}

%------------------------------------------------------------------------
% SECTION II.A: First order expansion in Delta
%------------------------------------------------------------------------
\textit{\label{sec:first_order_in_pairing}First order expansion in \texorpdfstring{$\hat{\Delta}$}{Delta}.}---Keeping only the leading-order term in $\hat{\Delta}$, $F(z)$ from Eq.~\eqref{eq:f_neumann_series} is approximated by
\begin{align}
    F(z) \approx F^{(1,\Delta)} & =  {(z - \hat{h})}^{-1} \hat{\Delta} {(z + \hat{h}^*)}^{-1}       \nonumber         \\
                                & = [-z^2 + \hat{h}^2]^{-1} A(z) [z^2 + (\hat{h}^*)^2]^{-1}, \label{eq:f_first_order}
\end{align}
with $A(z) = (z^2 \hat{\Delta} + z(\hat{h} \hat{\Delta} - \hat{\Delta} \hat{h}^*) - \hat{h} \hat{\Delta} \hat{h}^*)$.
We further write $F^{(1,\Delta)}$ as a sum of the even (e) and odd (o) parts in $z$, by identifying
\begin{equation}
    \label{eq:fe_fo_first_order}
    \begin{split}
        F_{\text{e}}^{(1,\Delta)} & = [-z^2 + \hat{h}^2]^{-1} (z^2 \hat{\Delta} - \hat{h} \hat{\Delta} \hat{h}^*) [z^2 + (\hat{h}^*)^2]^{-1}, \\
        F_{\text{o}}^{(1,\Delta)} & = [-z^2 + \hat{h}^2]^{-1} z(\hat{h} \hat{\Delta} - \hat{\Delta} \hat{h}^*) [z^2 + (\hat{h}^*)^2]^{-1}.
    \end{split}
\end{equation}
Upon inspection of these results, it is straightforward to see that Eq.~\eqref{eq:odd_freq_cond_triola} is a \emph{sufficient} condition for finite odd-frequency pairing, thus recovering the previous result derived for bulk multiband superconductors~\cite{triola_role_2020}, but now shown to be valid for {\it any} superconducting system, including both bulk and hybrid structures.

%------------------------------------------------------------------------
% SECTION III: Full small delta expansion of the anomalous correlations
%------------------------------------------------------------------------
\textit{\label{sec:small_delta_expansion}Full small-\texorpdfstring{$\hat{\Delta}$}{Delta} expansion.}---Having established the sufficient condition for any odd-frequency pairing in Eq.~\eqref{eq:odd_freq_cond_triola}, we next show that this first-order condition is also {\it necessary} for odd-frequency pairing to appear at any order in $\hat{\Delta}$.
For this, it is useful to start from Eq.~\eqref{eq:f_neumann_series} and then use the decomposition in Eq.~\eqref{eq:fe_fo_first_order} to write $F(z)$ as
\begin{equation}
    \label{eq:f_neumann_series_expanded}
    \begin{split}
        F(z) & =  \sum_{k=0}^{\infty} \left[  \left(F_{\text{e}}^{(1,\Delta)} + F_{\text{o}}^{(1,\Delta)}\right)  \hat{\Delta}^\dagger \right]^k \\
             & \quad \cdot \left(F_{\text{e}}^{(1,\Delta)} + F_{\text{o}}^{(1,\Delta)}\right)\,.
    \end{split}
\end{equation}
Here it is clear that higher-order terms in $\hat{\Delta}$ are obtained through the recursive equations for $n\ge 1$
\begin{align}
    F_{\text{e}}^{(n+1,\Delta)} & = F_{\text{e}}^{(1,\Delta)} \hat{\Delta}^\dagger F_{\text{e}}^{(n,\Delta)} + F_{\text{o}}^{(1,\Delta)} \hat{\Delta}^\dagger F_{\text{o}}^{(n,\Delta)}\,, \label{eq:rec1} \\
    F_{\text{o}}^{(n+1,\Delta)} & = F_{\text{e}}^{(1,\Delta)} \hat{\Delta}^\dagger F_{\text{o}}^{(n,\Delta)} + F_{\text{o}}^{(1,\Delta)} \hat{\Delta}^\dagger F_{\text{e}}^{(n,\Delta)}\,. \label{eq:rec2}
\end{align}
Thus, higher-order odd-frequency contributions can appear only if $F_{\text{o}}^{(1,\Delta)} \neq 0$. Therefore, Eq.~\eqref{eq:odd_freq_cond_triola} is both a \textit{sufficient} and \textit{necessary} condition for enabling odd-frequency pairing.
Furthermore, Eq.~\eqref{eq:rec2} reveals that coexistence between finite even- and odd-frequency pairs plays a fundamental role in the total odd-frequency pair amplitude.

%------------------------------------------------------------------------
% SECTION IV: High-frequency expansion
%------------------------------------------------------------------------
\textit{\label{sec:high_freq_exp}High-frequency expansion.}---While
% we above establish that
Eq.~\eqref{eq:odd_freq_cond_triola} is both sufficient and necessary for odd-frequency pairing% in any superconducting system
, it fails to provide details of the structure of the odd-frequency pairing.
In particular, we need conditions under which individual elements of $F(z)$ in Eq.~\eqref{eq:ftau} are finite, when summed over all orders in $\hat{\Delta}$.
For this, we instead perform a high-frequency expansion of $F(z)$.
Assuming high frequencies, i.e. $ \lVert \hat{h} / z \rVert < 1$, we start from Eq.~\eqref{eq:f_neumann_series} and expand the remaining inverses in terms of Neumann series.
%  ${(z - \hat{h})}^{-1}   = 1/z \sum_{k=0}^\infty {( \hat{h}/z )}^k$.
% \begin{align}
%     {(z - \hat{h})}^{-1}   & = \frac{1}{z} \sum_{k=0}^\infty {\left( \frac{\hat{h}}{z}
%         \right)}^k,
%     \\
%     {(z + \hat{h}^*)}^{-1} & = \frac{1}{z} \sum_{k=0}^\infty {\left( -\frac{\hat{h}^*}{z}
%         \right)}^k.
% \end{align}
Grouping terms of the same order in $1/z$ gives
\begin{equation}
    F(z) = \sum_{n=0}^\infty \frac{1}{z^{n+2}} \sum_\Lambda \underbrace{\hat{h}^{k_1} \hat{\Delta} (-\hat{h}^*)^{l_1}\hat{\Delta}^\dagger}_{m \text{ times}} \hat{h}^{k} \hat{\Delta} (-\hat{h}^*)^{l}, \label{eq:f_high_freq_exp}
\end{equation}
where $\Lambda$ contains all possible combinations of non-negative integers satisfying $k_1+l_1 + \ldots + k_m+l_m+k+l+2m=n$, such that the indicated term appears $m$ times in the sum.
This high-frequency expansion converges in the region where $|z|$ exceeds the largest eigenvalue of $H_{\text{BdG}}$. Still, we can draw conclusions about the presence of odd-frequency pairing in the entire complex plane.
To show this, we note that the inverse of a matrix is equal to its adjugate divided by its determinant, implying that any anomalous Green's function element is of the form
% \begin{equation}
$F_{ij}(z) = P_{ij}(z)/\det(z - H_{\text{BdG}})$.
% \end{equation}
Here, $\det(z - H_{\text{BdG}})$ is a polynomial in $z$ of order $2N$ for a system with $N$ states in the normal-state basis, and $P_{ij}(z)$ is a polynomial of order at most $2N-2$.
Thus, $F_{ij}(z)$ is holomorphic throughout the complex plane, except at the poles given by the eigenvalues of $H_{\text{BdG}}$. The series in Eq.~\eqref{eq:f_high_freq_exp} may then be analytically continued to the rest of the complex plane away from the poles. In particular, if the expansion is zero for large frequencies, it is also zero everywhere.

The lowest-order high-frequency contribution to odd-frequency pairing is obtained for $n=1$ in Eq.~\eqref{eq:f_high_freq_exp} as ${F_{\text{o}}^{(1,z)} = [\hat{h},\hat{\Delta}]_* / z^3}$.
% \begin{equation}
%     F_{\text{o}}^{(1,z)} = \frac{\hat{h} \hat{\Delta} - \hat{\Delta} \hat{h}^*}{z^3} \equiv \frac{[\hat{h},\hat{\Delta}]_*}{z^3}\,. \label{eq:f_series_n1}
% \end{equation}
Since $z$ is a continuously varying parameter, we conclude that a non-zero matrix element ${[\hat{h},\hat{\Delta}]_{*,ij} \neq 0}$ implies a finite odd-frequency component in Eq.~\eqref{eq:f_high_freq_exp} with the same indices, which establishes Eq.~\eqref{eq:odd_freq_cond_triola} also as \textit{sufficient} for the presence of finite odd-frequency pairing in the individual matrix element $F_{ij}(z)$.
Next, including both $n=1$ and $n=3$, we get
\begin{equation}
    \label{eq:high_freq_approx}
    \begin{split}
        F_{\text{o}}^{(3,z)} & = \frac{[\hat{h},\hat{\Delta}]_*}{z^3} + \frac{1}{z^5} \Big\{(\hat{h}^2 + \hat{\Delta} \hat{\Delta}^\dagger) [\hat{h},\hat{\Delta}]_* \\
                             & + [\hat{h},\hat{\Delta}]_* ((\hat{h}^*)^2 + \hat{\Delta}^\dagger \hat{\Delta}) \Big\}\,,
    \end{split}
\end{equation}
showing how higher order terms in Eq.~\eqref{eq:f_high_freq_exp} give additional \emph{sufficient} conditions for finite odd-frequency matrix elements.
Since all higher-order terms are obtained through simple matrix multiplications, without the need to invert matrices, the high-frequency expansion offers a computationally efficient way to study odd-frequency pairing. To showcase the strength of having general conditions for odd-frequency pairing, we below apply the results to several examples.

\begin{figure}[!t]
    \centering
    \includegraphics[width=\columnwidth]{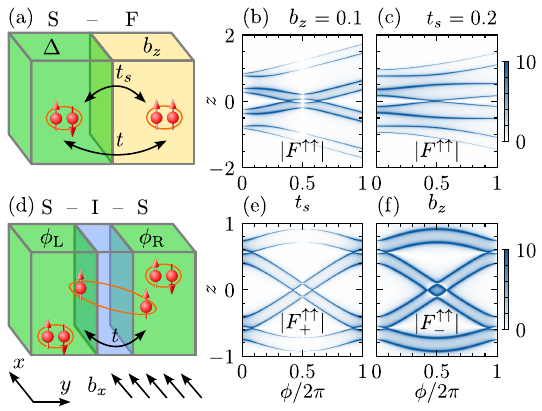}
    \caption{(a) SF junction with spin-preserving $t$ and spin-flip $t_s$ tunneling, order parameter $\Delta$ in S and spin-splitting $b_z$ in F. (b,c) Absolute value of $F_{\uparrow\uparrow}$ as a function of frequency $z$ and (b) $t_s$ or (c) $b_z$ for $\varepsilon_{\rm S} = 0.1$, $\varepsilon_{\rm F}=0.2$, $\Delta = 0.4$ and $t=0.5$. (d) Josephson junction with phases $\phi_L$ and $\phi_R$  across an insulator I in an in-plane magnetic field $b_x$. (e,f) Absolute value of odd-frequency $F_{+}^{\uparrow\uparrow}$ and even-frequency $F_{-}^{\uparrow\uparrow}$ pair amplitudes as a function of $z$ and $\phi = \phi_R - \phi_L$ across junction in (b) for $\varepsilon=0.1$, $t = 0.4 $, $\Delta=0.4$, and $b_x=0.1$.}
    \label{fig:junction}
\end{figure}

%------------------------------------------------------------------------
% SECTION V: Application to a SF junction
%------------------------------------------------------------------------
\textit{\label{sec:application0}Superconductor-ferromagnet junction.}---We start with the archetypal odd-frequency system: the superconductor-ferromagnet (SF) junction \cite{PhysRevLett.86.4096, RevModPhys.77.1321}, sketched in Fig.~\ref{fig:junction}(a).
Conceptually, it suffices to treat the S and F regions as point-like, with kinetic energies $\varepsilon_{\rm S,F}$, a magnetic field $b_z$ in F, and connected by a spin-active interface, with spin-conserving $t$ and spin-flip $t_s$ tunneling.
The normal-state Hamiltonian is given by
\begin{equation}
    \hat{h}       = \begin{pmatrix}
        \varepsilon_{\rm S} & 0                   & t                         & t_s                       \\
        0                   & \varepsilon_{\rm S} & t_s                       & t                         \\
        t                   & t_s                 & \varepsilon_{\rm F} + b_z & 0                         \\
        t_s                 & t                   & 0                         & \varepsilon_{\rm F} - b_z
    \end{pmatrix}\,, \label{eq:ham_sf}
\end{equation}
in the basis $\hat{\Phi}^T=\{c_{\rm S\uparrow}, c_{\rm S\downarrow}, c_{\rm F\uparrow}, c_{\rm F\downarrow}\}$. We further consider only spin-singlet pairing in the S region, such that only the matrix elements $\hat{\Delta}_{12} = -\hat{\Delta}_{21} = \Delta$ are non-zero.
Then, the condition for odd-frequency pairing in Eq.~\eqref{eq:odd_freq_cond_triola} results in
\begin{equation}
    \hat{h}\hat{\Delta} - \hat{\Delta}\hat{h}^* \neq 0 \quad\Leftrightarrow\quad t_s \Delta \neq 0\, \text{ or }\, t \Delta \neq 0.
\end{equation}
Thus, as long as $\Delta\neq0$, both the spin-flip tunneling $t_s$ and the regular spin-conserving tunneling $t$ are on their own enough to generate odd-frequency pairing.

We also examine equal spin-triplet correlations $F_{\uparrow\uparrow}(z)$ in the F region, which drive the long-range proximity effect \cite{RevModPhys.77.1321}.
Since the corresponding matrix element in Eq.~\eqref{eq:odd_freq_cond_triola} vanishes, we invoke the higher-order conditions of Eq.~\eqref{eq:f_high_freq_exp}, finding the first finite contribution at $n=5$:
\begin{equation}
    F_{\uparrow\uparrow}(z) \approx \frac{4 \Delta t\, t_s (t_s^2 - t^2)(b_z - \varepsilon_{\rm F} - \varepsilon_{\rm S})}{z^7}, \label{eq:sfjunction_fupup}
\end{equation}
which is notably odd in frequency. Thus, this calculation shows that odd-frequency equal-spin triplet $s$-wave pairing in the F region requires both types of tunneling $t$ and $t_s$, which in addition cannot have the same value. This is visualized in Figs.~\ref{fig:junction}(b,c), which show the absolute value of the full expression as a function of frequency $z$ and (b) $t_s$ and (c) $b_z$. However, a finite value of $b_z$ is not necessary, although it is needed to generate other spin-triplet states, see End Matter for full expressions.

%------------------------------------------------------------------------
% SECTION VI: Application to a Josephson junction under magnetic field
%------------------------------------------------------------------------
\textit{\label{sec:application2}Josephson junction in magnetic field.}---We next investigate the formation of odd-frequency pairing in a Josephson junction formed by two identical superconducting leads with a finite superconducting phase difference $\phi$ and subjected to a magnetic field $b_x$, see Fig.~\ref{fig:junction}(d). Here, we are interested in assessing the conditions for realizing odd-frequency pairing due to the interplay between interface effects, tunable by both $\phi$ and $b_x$. Within the same minimal description as above, the electron normal-state Hamiltonian is given by
\begin{equation}
    \hat{h}       = \begin{pmatrix}
        \varepsilon & b_x         & t           & 0           \\
        b_x         & \varepsilon & 0           & t           \\
        t           & 0           & \varepsilon & b_x         \\
        0           & t           & b_x         & \varepsilon
    \end{pmatrix}\,,
\end{equation}
written in the basis $\hat{\Phi}^T = \{c_{\rm L\uparrow}, c_{\rm L\downarrow}, c_{\rm R\uparrow}, c_{\rm R\downarrow}\}$, where $c_{i\sigma}$ annihilates an electron with spin $\sigma$ in the Left (Right) superconducting lead.
The superconducting part of the BdG Hamiltonian ($H_{\text{BdG}}$) is given by $\hat{\Delta}_{12} = -\hat{\Delta}_{21} = \Delta e^{i\phi/2}$ and $\hat{\Delta}_{34} = -\hat{\Delta}_{43} = \Delta e^{-i\phi/2}$,
with the superconducting phases $\phi_{\rm L} = \phi / 2$ and $\phi_{\rm R} = - \phi /2$.

To evaluate the conditions for odd-frequency pairing, we first apply Eq.~\eqref{eq:odd_freq_cond_triola}, resulting in
\begin{equation}
    \label{eq:ex2firstcond}
    \hat{h} \hat{\Delta} - \hat{\Delta} \hat{h}^* \neq 0\,\, \Leftrightarrow \,\,\Delta b_x \neq 0 \text{ or } t \Delta \sin(\phi/2) \neq 0\,.
\end{equation}
Thus, there are two sources of odd-frequency pairing. First, the applied magnetic field $b_x$ allows for odd-frequency pairing by breaking spin rotation symmetry \cite{RevModPhys.77.1321}, while the hopping $t$ breaks the continuous translational symmetry, thereby also allowing for odd-frequency pairing \cite{balatsky2018oddfreq,Cayao2020}.
Interestingly, the superconducting phase difference $\phi$ introduces a functional dependence that may be exploited for controlling odd-frequency Cooper pairs.
We next focus on induced equal-spin triplet pairing, relevant for superconducting spintronics \cite{Eschrig_2015, linder_superconducting_spintronics_2015, yang_superconducting_spintronics_2021,mel2022superconducting}.
It is the applied magnetic field $b_x$, coupling the two spin sectors of $H_{\text{BdG}}$, that leads to finite equal spin-triplet pairing across the junction, characterized by $F_{\rm LR}^{\uparrow \uparrow}$.
We isolate even and odd inter-lead pairing as $F_{\pm}^{\uparrow\uparrow} = \left(F_{\rm LR}^{\uparrow\uparrow} \pm F_{\rm RL}^{\uparrow\uparrow}\right)/2$, where
\begin{align}
    \label{eq:fplus}
    F_{+}^{\uparrow\uparrow}(z) & = \frac{8 b_x \varepsilon t \Delta z P \cos(\phi / 2)}{\mathrm{det}(z-H_{\text{BdG}})}\,, \\
    \label{eq:fminus}
    F_{-}^{\uparrow\uparrow}(z) & = \frac{-2 i b_x t \Delta Q(\phi)\sin(\phi/2)}{\mathrm{det}(z-H_{\text{BdG}})}\,,
\end{align}
with $P = \varepsilon^2 + t^2 + \Delta^2 - z^2 - b_x^2$, $Q(\phi) = \left(z^2 - b_x^2 + \Delta^2 + \varepsilon^2 + t^2 \right)^2 - 4 z^4 + 4 b_x^2 z^2 - 4 \varepsilon^2 t^2 - 4 \Delta^2 t^2 \sin^2 (\phi/2)$; here $\mathrm{det}(z-H_{\text{BdG}})$ is an even function in $z$.
Thus, $F_{+}^{\uparrow\uparrow}(z)$ has odd-frequency even-interlead spin-triplet symmetry, while $F_{-}^{\uparrow\uparrow}(z)$ has even-frequency odd-interlead spin-triplet symmetry; both consistent with the allowed pair symmetry classes \cite{PhysRevB.109.205406}.
The absolute value of both is plotted in Figs.~\ref{fig:junction}(e,f) as a function of $z$ and $\phi$ and clearly shows the appearance of Andreev bound states, with the applied magnetic field $b_x$ lifting the spin degeneracy. This demonstrates that $\phi$ can be used to control even- and odd-frequency pairings, where only the odd-(even-) frequency contribution remains at $\phi = 0$ ($\phi = \pi$).

We finally note that the phase dependence in Eq.~\eqref{eq:fplus} is different from Eq.~\eqref{eq:ex2firstcond}. This is due to the $\cos(\phi/2)$ term in Eq.~\eqref{eq:fplus} stemming from higher-order processes, which first appear at order $n=3$ in Eq.~\eqref{eq:f_high_freq_exp}, while the $\sin(\phi/2)$ term in Eq.~\eqref{eq:ex2firstcond} appears already at $n=1$ but then for odd-frequency inter-lead spin-singlet pairing. This reveals the importance of both having the general condition in Eq.~\eqref{eq:odd_freq_cond_triola} and extracting specific orbital terms through the high-frequency expansion in Eq.~\eqref{eq:f_high_freq_exp}.

%------------------------------------------------------------------------
% SECTION VII: Application to a proximitized TMD monolayer
%------------------------------------------------------------------------
%
\begin{figure}[!t]
    \centering
    \includegraphics[width=\columnwidth]{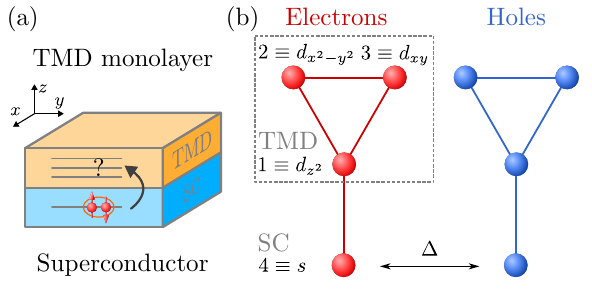}
    \caption{(a) TMD monolayer (orange) proximitized to a conventional superconductor (blue). (b) Electron (red) and hole (blue) orbitals with solid lines indicating the hopping. Orbitals $(1,2,3)$ model the TMD monolayer, and orbital four $(4)$ models the superconductor with order parameter $\Delta$.}
    \label{fig:heterostructure_levels}
\end{figure}

\textit{\label{sec:application1}Proximitized dichalcogenide monolayers.}---Finally, we study a transition metal dichalcogenide (TMD) monolayer proximitized by a conventional spin-singlet $s$-wave superconductor, see Fig.~\ref{fig:heterostructure_levels}(a).
TMD monolayers, such as MoS$_2$ or WSe$_2$, are accurately captured by three $d$-orbitals \cite{liu_three-band_2013}, sketched in Fig.~\ref{fig:heterostructure_levels}(b), where we then add a fourth orbital for the external superconductor. In the present discussion, the characteristic Ising spin-orbit coupling of the TMDs does not play a role because it does not couple the two spin sectors and it is thus sufficient to consider only one of them.
The electron normal-state Hamiltonian in momentum space is given by
\begin{equation}
    \label{eq:hamk_example1}
    \hat{h}_{\bm{k}} = \begin{pmatrix}
        H_{\bm{k},11} & H_{\bm{k},12} & H_{\bm{k},13} & H_{\bm{k},14} \\
        H_{\bm{k},21} & H_{\bm{k},22} & H_{\bm{k},23} & 0             \\
        H_{\bm{k},31} & H_{\bm{k},32} & H_{\bm{k},33} & 0             \\
        H_{\bm{k},41} & 0             & 0             & H_{\bm{k},44}
    \end{pmatrix},
\end{equation}
where $H_{\bm{k},ij}$ represents the finite matrix element between orbitals $i$ and $j$, see Fig.~\ref{fig:heterostructure_levels}(b). Only the first TMD orbital couples to the superconductor as it is the only orbital with substantial out-of-plane coupling.
Explicit forms of $H_{\bm{k},ij}$ are given in the End Matter.
Furthermore, a superconducting order parameter is only present in the superconductor, such that only $\hat{\Delta}^{44}_{\bm{k}} = \Delta$ is finite.

We proceed by evaluating Eq.~\eqref{eq:odd_freq_cond_triola}, giving
\begin{equation}
    \hat{h}_{\bm{k}} \hat{\Delta}_{\bm{k}} - \hat{\Delta}_{\bm{k}} \hat{h}_{-\bm{k}}^* \neq 0\quad \Leftrightarrow \quad \Delta H_{\bm{k},14} \neq 0\,.
\end{equation}
Thus, to generate any odd-frequency pairing, it is necessary and sufficient that $\Delta H_{\bm{k},14} \neq 0$, i.e.~as soon as the superconductor is actually superconducting and coupled to the TMD.
Interestingly, we find by analyzing the recursive Eqs.~(\ref{eq:rec1}-\ref{eq:rec2}), that both $F_{11}(\bm{k},z)$ and $F_{44}(\bm{k},z)$ never exhibit odd-frequency pairing, at any order.
We can apply the high-frequency expansion in Eq.~\eqref{eq:f_high_freq_exp} to answer two interesting questions this observation poses.
First, what would we need to break this apparent protection of even-frequency superconductivity?
From the high-frequency expansion at ${n=1}$, we find ${\Delta (H_{\bm{k},44} - H_{\bm{-k},44}) \neq 0}$ is needed for a finite odd-frequency $F_{44}(\bm{k},z)$ term, i.e.~the dispersion of the superconductor needs to break inversion symmetry.
In contrast, there is no contribution to $F_{11}(\bm{k},z)$ for $n=1$, as superconductivity in the TMD requires a higher-order process to reach orbital  $1$.
Evaluating the next-highest odd contribution at $n=3$, $F_{11}(\bm{k},z)$ exhibits an odd-frequency contribution if ${\Delta H_{\bm{k},14} H_{-\bm{k},41} (H_{-\bm{k},11} - H_{\bm{k},11} + H_{-\bm{k},44} - H_{\bm{k},44}) \neq 0}$.
This shows that either of the intra-orbital dispersions $H_{\bm{k},11}$ or $H_{\bm{k},44}$ needs to be break inversion symmetry for finite odd-frequency pairing.
Second, why do both $F_{22}(\bm{k},z)$ and $F_{33}(\bm{k},z)$ still contain odd-frequency contributions?
Again, applying Eq.~\eqref{eq:f_high_freq_exp}, now at order $n=5$, and after some simplifications, we find that this is caused by ${H_{-\bm{k},21} H_{\bm{k},23} H_{\bm{k},31} - H_{\bm{k},21} H_{-\bm{k},23} H_{-\bm{k},31} \neq 0}$, which is fulfilled because the TMD monolayer as a whole breaks inversion symmetry. Hence, the symmetry breaking of the normal state in the TMD is directly influencing the odd-frequency contributions.

%------------------------------------------------------------------------
% SECTION V: Conclusions
%------------------------------------------------------------------------
\textit{\label{sec:conclusions}Conclusions.}---We derive the necessary and sufficient conditions for odd-frequency superconducting pairing in any superconducting system, from bulk superconductors to superconducting hybrid structures. Further, a high-frequency expansion allows for computationally efficient analysis, giving direct access to individual components of the Cooper pair amplitude. As these results require only the BdG Hamiltonian, they provide a direct and fast way to both understand and engineer odd-frequency Cooper pairs. Ultimately, this predictive capability opens new avenues for tailoring the vast landscape of odd-frequency-driven quantum phenomena.

\textit{Acknowledgments.}---This work was supported by the Wallenberg Initiative Materials Science for Sustainability (WISE) funded by the Knut and Alice Wallenberg Foundation.
J. C. acknowledges  financial support from the Swedish Research Council  (Vetenskapsr\aa det Grant No.~2021-04121) and from the Olle Engkvist Foundation (Grant No.~243-1026).

%------------------------------------------------------------------------
% Bibliography 
\bibliography{library}

@article{PhysRevLett.116.257001,
  title     = {General Conditions for Proximity-Induced Odd-Frequency Superconductivity in Two-Dimensional Electronic Systems},
  author    = {Triola, Christopher and Badiane, Driss M. and Balatsky, Alexander V. and Rossi, E.},
  journal   = {Phys. Rev. Lett.},
  volume    = {116},
  issue     = {25},
  pages     = {257001},
  numpages  = {5},
  year      = {2016},
  month     = {Jun},
  publisher = {American Physical Society},
  doi       = {10.1103/PhysRevLett.116.257001},
  url       = {https://link.aps.org/doi/10.1103/PhysRevLett.116.257001}
}

@article{linder_odd-frequency_2019,
  title     = {Odd-frequency superconductivity},
  author    = {Linder, Jacob and Balatsky, Alexander V.},
  journal   = {Rev. Mod. Phys.},
  volume    = {91},
  issue     = {4},
  pages     = {045005},
  numpages  = {56},
  year      = {2019},
  month     = {Dec},
  publisher = {American Physical Society},
  doi       = {10.1103/RevModPhys.91.045005},
  url       = {https://link.aps.org/doi/10.1103/RevModPhys.91.045005}
}

@article{triola_role_2020,
  author   = {Triola, Christopher and Cayao, Jorge and Black-Schaffer, Annica M.},
  title    = {The Role of Odd-Frequency Pairing in Multiband Superconductors},
  journal  = {Ann. Phys.},
  volume   = {532},
  number   = {2},
  pages    = {1900298},
  doi      = {https://doi.org/10.1002/andp.201900298},
  url      = {https://onlinelibrary.wiley.com/doi/abs/10.1002/andp.201900298},
  year     = {2020}
}

@article{liu_three-band_2013,
  title     = {Three-band tight-binding model for monolayers of group-{VIB} transition metal dichalcogenides},
  author    = {Liu, Gui-Bin and Shan, Wen-Yu and Yao, Yugui and Yao, Wang and Xiao, Di},
  journal   = {Phys. Rev. B},
  volume    = {88},
  issue     = {8},
  pages     = {085433},
  numpages  = {10},
  year      = {2013},
  month     = {Aug},
  publisher = {American Physical Society},
  doi       = {10.1103/PhysRevB.88.085433},
  url       = {https://link.aps.org/doi/10.1103/PhysRevB.88.085433}
}

@article{PhysRevB.98.075425,
  title     = {Odd-frequency superconducting pairing in junctions with {R}ashba spin-orbit coupling},
  author    = {Cayao, Jorge and Black-Schaffer, Annica M.},
  journal   = {Phys. Rev. B},
  volume    = {98},
  issue     = {7},
  pages     = {075425},
  numpages  = {22},
  year      = {2018},
  month     = {Aug},
  publisher = {American Physical Society},
  doi       = {10.1103/PhysRevB.98.075425},
  url       = {https://link.aps.org/doi/10.1103/PhysRevB.98.075425}
}

@book{zagoskin,
  author    = {Alexandre Zagoskin},
  title     = {Quantum Theory of Many-Body Systems: Techniques and Applications},
  publisher = {Springer},
  year      = {2014}
}

@book{mahan2013many,
  title     = {Many-particle physics},
  author    = {Mahan, Gerald D},
  publisher = {Springer Science \& Business Media},
  year      = {2013}
}

@article{osti_4213537,
  title   = {New model of the anisotropic phase of superfluid {He$^3$}},
  author  = {Berezinskii, V. L.},
  journal = {Pis'ma Zh. Eksp. Teor. Fiz.},
  volume  = {20},
  issue   = {9},
  pages   = {628--631},
  year    = {1974},
  month   = {11},
  url     = {http://jetpletters.ru/ps/1792/article_27363.shtml}
}

@article{RevModPhys.77.1321,
  title     = {Odd triplet superconductivity and related phenomena in superconductor-ferromagnet structures},
  author    = {Bergeret, F. S. and Volkov, A. F. and Efetov, K. B.},
  journal   = {Rev. Mod. Phys.},
  volume    = {77},
  issue     = {4},
  pages     = {1321--1373},
  numpages  = {0},
  year      = {2005},
  month     = {Nov},
  publisher = {American Physical Society},
  doi       = {10.1103/RevModPhys.77.1321},
  url       = {https://link.aps.org/doi/10.1103/RevModPhys.77.1321}
}

@article{petrashov1994conductivity,
  title   = {Conductivity of mesoscopic structures with ferromagnetic and superconducting regions},
  author  = {Petrashov, VT and Antonov, VN and Maksimov, SV and Shaikhaidarov, R Sh},
  journal = {JETP Letters},
  volume  = {59},
  number  = {8},
  pages   = {551--555},
  year    = {1994},
  url     = {http://jetpletters.ru/ps/0/article_19787.shtml}
}

@article{A.Kadigrobov_2001,
  doi       = {10.1209/epl/i2001-00107-2},
  url       = {https://doi.org/10.1209/epl/i2001-00107-2},
  year      = {2001},
  month     = {may},
  publisher = {},
  volume    = {54},
  number    = {3},
  pages     = {394},
  author    = {A. Kadigrobov and R. I. Shekhter and M. Jonson},
  title     = {Quantum spin fluctuations as a source of long-range 
               proximity effects in diffusive ferromagnet-super conductor
               structures},
  journal   = {EPL}
}

@article{PhysRevLett.86.4096,
  title     = {Long-Range Proximity Effects in Superconductor-Ferromagnet Structures},
  author    = {Bergeret, F. S. and Volkov, A. F. and Efetov, K. B.},
  journal   = {Phys. Rev. Lett.},
  volume    = {86},
  issue     = {18},
  pages     = {4096--4099},
  numpages  = {0},
  year      = {2001},
  month     = {Apr},
  publisher = {American Physical Society},
  doi       = {10.1103/PhysRevLett.86.4096},
  url       = {https://link.aps.org/doi/10.1103/PhysRevLett.86.4096}
}

@article{PhysRevLett.98.077003,
  title     = {Fully Developed Triplet Proximity Effect},
  author    = {Braude, V. and Nazarov, Yu. V.},
  journal   = {Phys. Rev. Lett.},
  volume    = {98},
  issue     = {7},
  pages     = {077003},
  numpages  = {4},
  year      = {2007},
  month     = {Feb},
  publisher = {American Physical Society},
  doi       = {10.1103/PhysRevLett.98.077003},
  url       = {https://link.aps.org/doi/10.1103/PhysRevLett.98.077003}
}

@article{PhysRevLett.99.127002,
  title     = {Odd Triplet Pairing in Clean Superconductor/Ferromagnet Heterostructures},
  author    = {Halterman, Klaus and Barsic, Paul H. and Valls, Oriol T.},
  journal   = {Phys. Rev. Lett.},
  volume    = {99},
  issue     = {12},
  pages     = {127002},
  numpages  = {4},
  year      = {2007},
  month     = {Sep},
  publisher = {American Physical Society},
  doi       = {10.1103/PhysRevLett.99.127002},
  url       = {https://link.aps.org/doi/10.1103/PhysRevLett.99.127002}
}

@article{PhysRevB.88.104514,
  title     = {Odd-frequency superconducting pairing in multiband superconductors},
  author    = {Black-Schaffer, Annica M. and Balatsky, Alexander V.},
  journal   = {Phys. Rev. B},
  volume    = {88},
  issue     = {10},
  pages     = {104514},
  numpages  = {5},
  year      = {2013},
  month     = {Sep},
  publisher = {American Physical Society},
  doi       = {10.1103/PhysRevB.88.104514},
  url       = {https://link.aps.org/doi/10.1103/PhysRevB.88.104514}
}

@article{PhysRevB.93.201402,
  title     = {All-electrical generation and control of odd-frequency $s$-wave {Cooper} pairs in double quantum dots},
  author    = {Burset, Pablo and Lu, Bo and Ebisu, Hiromi and Asano, Yasuhiro and Tanaka, Yukio},
  journal   = {Phys. Rev. B},
  volume    = {93},
  issue     = {20},
  pages     = {201402},
  numpages  = {6},
  year      = {2016},
  month     = {May},
  publisher = {American Physical Society},
  doi       = {10.1103/PhysRevB.93.201402},
  url       = {https://link.aps.org/doi/10.1103/PhysRevB.93.201402}
}

@article{PhysRevB.97.214508,
  title     = {Green's-function theory of dirty two-band superconductivity},
  author    = {Asano, Yasuhiro and Golubov, Alexander A.},
  journal   = {Phys. Rev. B},
  volume    = {97},
  issue     = {21},
  pages     = {214508},
  numpages  = {15},
  year      = {2018},
  month     = {Jun},
  publisher = {American Physical Society},
  doi       = {10.1103/PhysRevB.97.214508},
  url       = {https://link.aps.org/doi/10.1103/PhysRevB.97.214508}
}

@article{PhysRevB.92.094517,
  title     = {Experimentally observable signatures of odd-frequency pairing in multiband superconductors},
  author    = {Komendov\'a, L. and Balatsky, A. V. and Black-Schaffer, A. M.},
  journal   = {Phys. Rev. B},
  volume    = {92},
  issue     = {9},
  pages     = {094517},
  numpages  = {7},
  year      = {2015},
  month     = {Sep},
  publisher = {American Physical Society},
  doi       = {10.1103/PhysRevB.92.094517},
  url       = {https://link.aps.org/doi/10.1103/PhysRevB.92.094517}
}

@article{PhysRevLett.119.087001,
  title     = {Odd-Frequency Superconductivity in {Sr$_2$RuO$_4$} Measured by {K}err Rotation},
  author    = {Komendov\'a, L. and Black-Schaffer, A. M.},
  journal   = {Phys. Rev. Lett.},
  volume    = {119},
  issue     = {8},
  pages     = {087001},
  numpages  = {5},
  year      = {2017},
  month     = {Aug},
  publisher = {American Physical Society},
  doi       = {10.1103/PhysRevLett.119.087001},
  url       = {https://link.aps.org/doi/10.1103/PhysRevLett.119.087001}
}

@article{mel2022superconducting,
  title   = {Superconducting spintronics: state of the art and prospects},
  author  = {Mel'nikov, AS and Mironov, Sergei Viktorovich and Samokhvalov, Aleksei Vladimirovich and Buzdin, Alexander Ivanovich},
  journal = {Uspekhi Fiz. Nauk},
  volume  = {192},
  pages   = {1339--1384},
  url     = {https://doi.org/10.3367/UFNe.2021.07.039020},
  year    = {2022}
}

@article{yang_superconducting_spintronics_2021,
  author  = {Yang, Guang and Ciccarelli, Chiara and Robinson, Jason W. A.},
  title   = {Boosting spintronics with superconductivity},
  journal = {APL Mater.},
  volume  = {9},
  number  = {5},
  pages   = {050703},
  year    = {2021},
  month   = {05},
  issn    = {2166-532X},
  doi     = {10.1063/5.0048904},
  url     = {https://doi.org/10.1063/5.0048904}
}

@article{linder_superconducting_spintronics_2015,
  author  = {Linder, Jacob
             and Robinson, Jason W. A.},
  title   = {Superconducting spintronics},
  journal = {Nat. Phys.},
  year    = {2015},
  month   = {Apr},
  day     = {01},
  volume  = {11},
  number  = {4},
  pages   = {307-315},
  issn    = {1745-2481},
  doi     = {10.1038/nphys3242},
  url     = {https://doi.org/10.1038/nphys3242}
}

@article{ramires_fitness_2018,
  title     = {Tailoring ${T}_{c}$ by symmetry principles: The concept of superconducting fitness},
  author    = {Ramires, Aline and Agterberg, Daniel F. and Sigrist, Manfred},
  journal   = {Phys. Rev. B},
  volume    = {98},
  issue     = {2},
  pages     = {024501},
  numpages  = {6},
  year      = {2018},
  month     = {Jul},
  publisher = {American Physical Society},
  doi       = {10.1103/PhysRevB.98.024501},
  url       = {https://link.aps.org/doi/10.1103/PhysRevB.98.024501}
}

@article{PhysRevB.68.064513,
  title     = {Manifestation of triplet superconductivity in superconductor-ferromagnet structures},
  author    = {Bergeret, F. S. and Volkov, A. F. and Efetov, K. B.},
  journal   = {Phys. Rev. B},
  volume    = {68},
  issue     = {6},
  pages     = {064513},
  numpages  = {11},
  year      = {2003},
  month     = {Aug},
  publisher = {American Physical Society},
  doi       = {10.1103/PhysRevB.68.064513},
  url       = {https://link.aps.org/doi/10.1103/PhysRevB.68.064513}
}

@article{RevModPhys.77.935,
  title     = {Proximity effects in superconductor-ferromagnet heterostructures},
  author    = {Buzdin, A. I.},
  journal   = {Rev. Mod. Phys.},
  volume    = {77},
  issue     = {3},
  pages     = {935--976},
  numpages  = {0},
  year      = {2005},
  month     = {Sep},
  publisher = {American Physical Society},
  doi       = {10.1103/RevModPhys.77.935},
  url       = {https://link.aps.org/doi/10.1103/RevModPhys.77.935}
}

@article{PhysRevB.72.140503,
  title     = {Anomalous features of the proximity effect in triplet superconductors},
  author    = {Tanaka, Y. and Asano, Y. and Golubov, A. A. and Kashiwaya, S.},
  journal   = {Phys. Rev. B},
  volume    = {72},
  issue     = {14},
  pages     = {140503},
  numpages  = {4},
  year      = {2005},
  month     = {Oct},
  publisher = {American Physical Society},
  doi       = {10.1103/PhysRevB.72.140503},
  url       = {https://link.aps.org/doi/10.1103/PhysRevB.72.140503}
}

@article{Eschrig_2015,
  doi       = {10.1088/0034-4885/78/10/104501},
  url       = {https://doi.org/10.1088/0034-4885/78/10/104501},
  year      = {2015},
  month     = {sep},
  publisher = {IOP Publishing},
  volume    = {78},
  number    = {10},
  pages     = {104501},
  author    = {Eschrig, Matthias},
  title     = {Spin-polarized supercurrents for spintronics: a review of current progress},
  journal   = {Rep. Prog. Phys.}
}

@article{PhysRevB.92.054516,
  title     = {Ab initio theory of magnetic-field-induced odd-frequency two-band superconductivity in {MgB}$_2$},
  author    = {Aperis, Alex and Maldonado, Pablo and Oppeneer, Peter M.},
  journal   = {Phys. Rev. B},
  volume    = {92},
  issue     = {5},
  pages     = {054516},
  numpages  = {8},
  year      = {2015},
  month     = {Aug},
  publisher = {American Physical Society},
  doi       = {10.1103/PhysRevB.92.054516},
  url       = {https://link.aps.org/doi/10.1103/PhysRevB.92.054516}
}

@article{Keizer2006,
  author  = {Keizer, R. S.
             and Goennenwein, S. T. B.
             and Klapwijk, T. M.
             and Miao, G.
             and Xiao, G.
             and Gupta, A.},
  title   = {A spin triplet supercurrent through the half-metallic ferromagnet {CrO}$_2$},
  journal = {Nature},
  year    = {2006},
  month   = {Feb},
  day     = {01},
  volume  = {439},
  number  = {7078},
  pages   = {825-827},
  issn    = {1476-4687},
  doi     = {10.1038/nature04499},
  url     = {https://doi.org/10.1038/nature04499}
}

@article{PhysRevLett.104.137002,
  title     = {Observation of Spin-Triplet Superconductivity in {Co}-Based {J}osephson Junctions},
  author    = {Khaire, Trupti S. and Khasawneh, Mazin A. and Pratt, W. P. and Birge, Norman O.},
  journal   = {Phys. Rev. Lett.},
  volume    = {104},
  issue     = {13},
  pages     = {137002},
  numpages  = {4},
  year      = {2010},
  month     = {Mar},
  publisher = {American Physical Society},
  doi       = {10.1103/PhysRevLett.104.137002},
  url       = {https://link.aps.org/doi/10.1103/PhysRevLett.104.137002}
}

@article{science.1189246,
  author  = {J. W. A. Robinson  and J. D. S. Witt  and M. G. Blamire },
  title   = {Controlled Injection of Spin-Triplet Supercurrents into a Strong Ferromagnet},
  journal = {Science},
  volume  = {329},
  number  = {5987},
  pages   = {59-61},
  year    = {2010},
  doi     = {10.1126/science.1189246},
  url     = {https://www.science.org/doi/abs/10.1126/science.1189246}
}

@article{PhysRevX.5.041021,
  title     = {Intrinsic Paramagnetic {M}eissner Effect Due to $s$-Wave Odd-Frequency Superconductivity},
  author    = {Di Bernardo, A. and Salman, Z. and Wang, X. L. and Amado, M. and Egilmez, M. and Flokstra, M. G. and Suter, A. and Lee, S. L. and Zhao, J. H. and Prokscha, T. and Morenzoni, E. and Blamire, M. G. and Linder, J. and Robinson, J. W. A.},
  journal   = {Phys. Rev. X},
  volume    = {5},
  issue     = {4},
  pages     = {041021},
  numpages  = {7},
  year      = {2015},
  month     = {Nov},
  publisher = {American Physical Society},
  doi       = {10.1103/PhysRevX.5.041021},
  url       = {https://link.aps.org/doi/10.1103/PhysRevX.5.041021}
}

@article{PhysRevB.95.184506,
  title     = {Odd-frequency superconductivity in a nanowire coupled to {M}ajorana zero modes},
  author    = {Lee, Shu-Ping and Lutchyn, Roman M. and Maciejko, Joseph},
  journal   = {Phys. Rev. B},
  volume    = {95},
  issue     = {18},
  pages     = {184506},
  numpages  = {12},
  year      = {2017},
  month     = {May},
  publisher = {American Physical Society},
  doi       = {10.1103/PhysRevB.95.184506},
  url       = {https://link.aps.org/doi/10.1103/PhysRevB.95.184506}
}

@article{JPSJ.81.011013,
  author  = {Tanaka ,Yukio and Sato ,Masatoshi and Nagaosa ,Naoto},
  title   = {Symmetry and Topology in Superconductors –Odd-Frequency Pairing and Edge States–},
  journal = {J. Phys. Soc. Jpn.},
  volume  = {81},
  number  = {1},
  pages   = {011013},
  year    = {2012},
  doi     = {10.1143/JPSJ.81.011013},
  url     = {https://doi.org/10.1143/JPSJ.81.011013}
}

@article{Cayao2020,
  author  = {Cayao, Jorge
             and Triola, Christopher
             and Black-Schaffer, Annica M.},
  title   = {Odd-frequency superconducting pairing in one-dimensional systems},
  journal = {Eur. Phys. J.: Spec. Top.},
  year    = {2020},
  month   = {Feb},
  day     = {01},
  volume  = {229},
  number  = {4},
  pages   = {545-575},
  issn    = {1951-6401},
  doi     = {10.1140/epjst/e2019-900168-0},
  url     = {https://doi.org/10.1140/epjst/e2019-900168-0}
}

@article{PhysRevB.52.1271,
  title     = {Properties of odd-gap superconductors},
  author    = {Abrahams, Elihu and Balatsky, Alexander and Scalapino, D. J. and Schrieffer, J. R.},
  journal   = {Phys. Rev. B},
  volume    = {52},
  issue     = {2},
  pages     = {1271--1278},
  numpages  = {0},
  year      = {1995},
  month     = {Jul},
  publisher = {American Physical Society},
  doi       = {10.1103/PhysRevB.52.1271},
  url       = {https://link.aps.org/doi/10.1103/PhysRevB.52.1271}
}

@article{PhysRevB.89.184508,
  title     = {Paramagnetic instability of small topological superconductors},
  author    = {Suzuki, Shu-Ichiro and Asano, Yasuhiro},
  journal   = {Phys. Rev. B},
  volume    = {89},
  issue     = {18},
  pages     = {184508},
  numpages  = {7},
  year      = {2014},
  month     = {May},
  publisher = {American Physical Society},
  doi       = {10.1103/PhysRevB.89.184508},
  url       = {https://link.aps.org/doi/10.1103/PhysRevB.89.184508}
}

@article{PhysRevLett.125.117003,
  title     = {Unveiling Odd-Frequency Pairing around a Magnetic Impurity in a Superconductor},
  author    = {Perrin, Vivien and Santos, Fl\'avio L. N. and M\'enard, Gerbold C. and Brun, Christophe and Cren, Tristan and Civelli, Marcello and Simon, Pascal},
  journal   = {Phys. Rev. Lett.},
  volume    = {125},
  issue     = {11},
  pages     = {117003},
  numpages  = {6},
  year      = {2020},
  month     = {Sep},
  publisher = {American Physical Society},
  doi       = {10.1103/PhysRevLett.125.117003},
  url       = {https://link.aps.org/doi/10.1103/PhysRevLett.125.117003}
}

@article{PhysRevMaterials.5.114801,
  title     = {Unconventional {M}eissner screening induced by chiral molecules in a conventional superconductor},
  author    = {Alpern, Hen and Amundsen, Morten and Hartmann, Roman and Sukenik, Nir and Spuri, Alfredo and Yochelis, Shira and Prokscha, Thomas and Gutkin, Vitaly and Anahory, Yonathan and Scheer, Elke and Linder, Jacob and Salman, Zaher and Millo, Oded and Paltiel, Yossi and Di Bernardo, Angelo},
  journal   = {Phys. Rev. Mater.},
  volume    = {5},
  issue     = {11},
  pages     = {114801},
  numpages  = {17},
  year      = {2021},
  month     = {Nov},
  publisher = {American Physical Society},
  doi       = {10.1103/PhysRevMaterials.5.114801},
  url       = {https://link.aps.org/doi/10.1103/PhysRevMaterials.5.114801}
}

@article{c57s-skv9,
  title     = {Diamagnetic {M}eissner response of odd-frequency superconducting pairing from quantum geometry},
  author    = {Bhattacharya, Ankita and Black-Schaffer, Annica M.},
  journal   = {Phys. Rev. B},
  volume    = {113},
  issue     = {9},
  pages     = {094501},
  numpages  = {11},
  year      = {2026},
  month     = {Mar},
  publisher = {American Physical Society},
  doi       = {10.1103/c57s-skv9},
  url       = {https://link.aps.org/doi/10.1103/c57s-skv9}
}

@article{PhysRevB.95.174516,
  title     = {Majorana {STM} as a perfect detector of odd-frequency superconductivity},
  author    = {Kashuba, Oleksiy and Sothmann, Bj\"orn and Burset, Pablo and Trauzettel, Bj\"orn},
  journal   = {Phys. Rev. B},
  volume    = {95},
  issue     = {17},
  pages     = {174516},
  numpages  = {9},
  year      = {2017},
  month     = {May},
  publisher = {American Physical Society},
  doi       = {10.1103/PhysRevB.95.174516},
  url       = {https://link.aps.org/doi/10.1103/PhysRevB.95.174516}
}

@article{PhysRevB.94.104501,
  title     = {Identifying detrimental effects for multiorbital superconductivity: Application to {Sr}$_2${RuO}$_4$},
  author    = {Ramires, Aline and Sigrist, Manfred},
  journal   = {Phys. Rev. B},
  volume    = {94},
  issue     = {10},
  pages     = {104501},
  numpages  = {13},
  year      = {2016},
  month     = {Sep},
  publisher = {American Physical Society},
  doi       = {10.1103/PhysRevB.94.104501},
  url       = {https://link.aps.org/doi/10.1103/PhysRevB.94.104501}
}

@article{Fischer_2013,
  doi       = {10.1088/1367-2630/15/7/073006},
  url       = {https://doi.org/10.1088/1367-2630/15/7/073006},
  year      = {2013},
  month     = {jul},
  publisher = {IOP Publishing},
  volume    = {15},
  number    = {7},
  pages     = {073006},
  author    = {Fischer, Mark H},
  title     = {Gap symmetry and stability analysis in the multi-orbital {Fe}-based superconductors},
  journal   = {New. J. Phys.}
}

@article{kayatz2026tmd,
  title   = {Superconducting properties of transition metal dichalcogenides in proximity to a conventional superconductor},
  author  = {Florian Kayatz and Annica M. Black-Schaffer and Jorge Cayao},
  journal = {arXiv:2601.21994},
  url     = {https://arxiv.org/abs/2601.21994},
  year    = {2026}
}

@article{10.1093/ptep/ptae065,
  author  = {Tanaka, Yukio and Tamura, Shun and Cayao, Jorge},
  title   = {Theory of {Majorana} Zero Modes in Unconventional Superconductors},
  journal = {Prog. Theor. Exp. Phys.},
  volume  = {2024},
  number  = {8},
  pages   = {08C105},
  year    = {2024},
  month   = {08},
  issn    = {2050-3911},
  doi     = {10.1093/ptep/ptae065},
  url     = {https://doi.org/10.1093/ptep/ptae065}
}

@article{c325-kgbf,
  title     = {Detecting the topological phase transition in superconductor-semiconductor hybrids by electronic {R}aman spectroscopy},
  author    = {Mizushima, Takeshi and Tanaka, Yukio and Cayao, Jorge},
  journal   = {Phys. Rev. B},
  volume    = {112},
  issue     = {17},
  pages     = {174504},
  numpages  = {21},
  year      = {2025},
  month     = {Nov},
  publisher = {American Physical Society},
  doi       = {10.1103/c325-kgbf},
  url       = {https://link.aps.org/doi/10.1103/c325-kgbf}
}

@article{PhysRevB.111.024507,
  title     = {Odd-frequency superconducting pairing due to multiple {M}ajorana edge modes in driven topological superconductors},
  author    = {Ahmed, Eslam and Tamura, Shun and Tanaka, Yukio and Cayao, Jorge},
  journal   = {Phys. Rev. B},
  volume    = {111},
  issue     = {2},
  pages     = {024507},
  numpages  = {17},
  year      = {2025},
  month     = {Jan},
  publisher = {American Physical Society},
  doi       = {10.1103/PhysRevB.111.024507},
  url       = {https://link.aps.org/doi/10.1103/PhysRevB.111.024507}
}

@article{ahmed2025anomalous,
  author  = {Ahmed, Eslam
             and Tanaka, Yukio
             and Cayao, Jorge},
  title   = {Anomalous Proximity Effect Under {A}ndreev and {M}ajorana Bound States},
  journal = {J. Supercond. Nov. Magn.},
  year    = {2025},
  month   = {Oct},
  day     = {08},
  volume  = {38},
  number  = {5},
  pages   = {220},
  issn    = {1557-1947},
  doi     = {10.1007/s10948-025-07057-9},
  url     = {https://doi.org/10.1007/s10948-025-07057-9}
}

@article{chakraborty2026breathing,
  title   = {Breathing mode inducing dynamical pairing in {Kagome} materials},
  author  = {Debmalya Chakraborty and Anushree Datta and Jorge Cayao},
  journal = {arXiv:2607.01052},
  url     = {https://arxiv.org/abs/2607.01052},
  year    = {2026}
}

@article{sardinero2026odd,
  title     = {Odd-Frequency Pairing in {J}osephson Junctions Coupled by Magnetic Textures},
  author    = {Sardinero, Ignacio and Cayao, Jorge and Seoane Souto, Rub{\'e}n and Burset, Pablo},
  journal   = {Phys. Status Solidi RRL},
  volume    = {20},
  number    = {1},
  pages     = {e202500413},
  year      = {2026},
  doi       = {https://doi.org/10.1002/pssr.202500413},
  publisher = {Wiley Online Library}
}

@article{fksg-x8pr,
  title     = {Odd-frequency pairing due to {M}ajorana and trivial {A}ndreev bound states},
  author    = {Ahmed, Eslam and Tamura, Shun and Tanaka, Yukio and Cayao, Jorge},
  journal   = {Phys. Rev. B},
  volume    = {111},
  issue     = {22},
  pages     = {224508},
  numpages  = {17},
  year      = {2025},
  month     = {Jun},
  publisher = {American Physical Society},
  doi       = {10.1103/fksg-x8pr},
  url       = {https://link.aps.org/doi/10.1103/fksg-x8pr}
}

@article{PhysRevB.110.125408,
  title     = {Emergent pair symmetries in systems with poor man's {M}ajorana modes},
  author    = {Cayao, Jorge},
  journal   = {Phys. Rev. B},
  volume    = {110},
  issue     = {12},
  pages     = {125408},
  numpages  = {10},
  year      = {2024},
  month     = {Sep},
  publisher = {American Physical Society},
  doi       = {10.1103/PhysRevB.110.125408},
  url       = {https://link.aps.org/doi/10.1103/PhysRevB.110.125408}
}

@article{PhysRevB.96.155426,
  title     = {Odd-frequency superconducting pairing and subgap density of states at the edge of a two-dimensional topological insulator without magnetism},
  author    = {Cayao, Jorge and Black-Schaffer, Annica M.},
  journal   = {Phys. Rev. B},
  volume    = {96},
  issue     = {15},
  pages     = {155426},
  numpages  = {22},
  year      = {2017},
  month     = {Oct},
  publisher = {American Physical Society},
  doi       = {10.1103/PhysRevB.96.155426},
  url       = {https://link.aps.org/doi/10.1103/PhysRevB.96.155426}
}

@article{PhysRevB.92.134512,
  title     = {Proximity-induced triplet superconductivity in {R}ashba materials},
  author    = {Reeg, Christopher R. and Maslov, Dmitrii L.},
  journal   = {Phys. Rev. B},
  volume    = {92},
  issue     = {13},
  pages     = {134512},
  numpages  = {10},
  year      = {2015},
  month     = {Oct},
  publisher = {American Physical Society},
  doi       = {10.1103/PhysRevB.92.134512},
  url       = {https://link.aps.org/doi/10.1103/PhysRevB.92.134512}
}

@article{FukayaJPCM2025,
  author  = {Fukaya, Yuri and Lu, Bo and Yada, Keiji and Tanaka, Yukio and Cayao, Jorge},
  title   = {Superconducting phenomena in systems with unconventional magnets},
  journal = {J. Phys.: Condens. Matter},
  volume  = {37},
  pages   = {313003},
  year    = {2025},
  url     = {https://iopscience.iop.org/article/10.1088/1361-648X/adf1cf}
}

@article{PhysRevB.92.100507,
  title     = {Odd-frequency triplet superconductivity at the helical edge of a topological insulator},
  author    = {Cr\'epin, Francois and Burset, Pablo and Trauzettel, Bj\"orn},
  journal   = {Phys. Rev. B},
  volume    = {92},
  issue     = {10},
  pages     = {100507(R)},
  numpages  = {5},
  year      = {2015},
  month     = {Sep},
  publisher = {American Physical Society},
  doi       = {10.1103/PhysRevB.92.100507},
  url       = {https://link.aps.org/doi/10.1103/PhysRevB.92.100507}
}

@article{PhysRevB.92.205424,
  title     = {Superconducting proximity effect in three-dimensional topological insulators in the presence of a magnetic field},
  author    = {Burset, Pablo and Lu, Bo and Tkachov, Grigory and Tanaka, Yukio and Hankiewicz, Ewelina M. and Trauzettel, Bj\"orn},
  journal   = {Phys. Rev. B},
  volume    = {92},
  issue     = {20},
  pages     = {205424},
  numpages  = {14},
  year      = {2015},
  month     = {Nov},
  publisher = {American Physical Society},
  doi       = {10.1103/PhysRevB.92.205424},
  url       = {https://link.aps.org/doi/10.1103/PhysRevB.92.205424}
}

@article{PhysRevB.90.220501,
  title     = {Unconventional superconductivity in double quantum dots},
  author    = {Sothmann, Bj\"orn and Weiss, Stephan and Governale, Michele and K\"onig, J\"urgen},
  journal   = {Phys. Rev. B},
  volume    = {90},
  issue     = {22},
  pages     = {220501(R)},
  numpages  = {5},
  year      = {2014},
  month     = {Dec},
  publisher = {American Physical Society},
  doi       = {10.1103/PhysRevB.90.220501},
  url       = {https://link.aps.org/doi/10.1103/PhysRevB.90.220501}
}

@article{PhysRevLett.120.037701,
  title     = {Creation of Spin-Triplet {C}ooper Pairs in the Absence of Magnetic Ordering},
  author    = {Breunig, Daniel and Burset, Pablo and Trauzettel, Bj\"orn},
  journal   = {Phys. Rev. Lett.},
  volume    = {120},
  issue     = {3},
  pages     = {037701},
  numpages  = {6},
  year      = {2018},
  month     = {Jan},
  publisher = {American Physical Society},
  doi       = {10.1103/PhysRevLett.120.037701},
  url       = {https://link.aps.org/doi/10.1103/PhysRevLett.120.037701}
}

@article{PhysRevB.111.064502,
  title     = {Josephson effect and odd-frequency pairing in superconducting junctions with unconventional magnets},
  author    = {Fukaya, Yuri and Maeda, Kazuki and Yada, Keiji and Cayao, Jorge and Tanaka, Yukio and Lu, Bo},
  journal   = {Phys. Rev. B},
  volume    = {111},
  issue     = {6},
  pages     = {064502},
  numpages  = {15},
  year      = {2025},
  month     = {Feb},
  publisher = {American Physical Society},
  doi       = {10.1103/PhysRevB.111.064502},
  url       = {https://link.aps.org/doi/10.1103/PhysRevB.111.064502}
}

@article{Maeda2025,
  title     = {Classification of pair symmetries in superconductors with unconventional magnetism},
  author    = {Maeda, Kazuki and Fukaya, Yuri and Yada, Keiji and Lu, Bo and Tanaka, Yukio and Cayao, Jorge},
  journal   = {Phys. Rev. B},
  volume    = {111},
  issue     = {14},
  pages     = {144508},
  numpages  = {14},
  year      = {2025},
  month     = {Apr},
  publisher = {American Physical Society},
  doi       = {10.1103/PhysRevB.111.144508},
  url       = {https://link.aps.org/doi/10.1103/PhysRevB.111.144508}
}

@article{PhysRevB.103.104505,
  title     = {Floquet engineering bulk odd-frequency superconducting pairs},
  author    = {Cayao, Jorge and Triola, Christopher and Black-Schaffer, Annica M.},
  journal   = {Phys. Rev. B},
  volume    = {103},
  issue     = {10},
  pages     = {104505},
  numpages  = {6},
  year      = {2021},
  month     = {Mar},
  publisher = {American Physical Society},
  doi       = {10.1103/PhysRevB.103.104505},
  url       = {https://link.aps.org/doi/10.1103/PhysRevB.103.104505}
}

@article{PhysRevB.109.134517,
  title     = {Floquet engineering {H}iggs dynamics in time-periodic superconductors},
  author    = {Kuhn, Tobias and Sothmann, Bj\"orn and Cayao, Jorge},
  journal   = {Phys. Rev. B},
  volume    = {109},
  issue     = {13},
  pages     = {134517},
  numpages  = {11},
  year      = {2024},
  month     = {Apr},
  publisher = {American Physical Society},
  doi       = {10.1103/PhysRevB.109.134517},
  url       = {https://link.aps.org/doi/10.1103/PhysRevB.109.134517}
}

@article{SciPostPhys.20.2.059,
  title     = {Light-induced {F}loquet spin-triplet {Cooper} pairs in unconventional magnets},
  pages     = {059},
  author    = {Fu, Pei-Hao and Mondal, Sayan and Liu, Jun-Feng and Cayao, Jorge},
  journal   = {SciPost Phys.},
  volume    = {20},
  year      = {2026},
  publisher = {SciPost},
  doi       = {10.21468/SciPostPhys.20.2.059},
  url       = {https://scipost.org/10.21468/SciPostPhys.20.2.059}
}

@article{lkf9-jgv6,
  title     = {Floquet Engineering Spin Triplet States in Unconventional Magnets},
  author    = {Fu, Pei-Hao and Mondal, Sayan and Liu, Jun-Feng and Tanaka, Yukio and Cayao, Jorge},
  journal   = {Phys. Rev. Lett.},
  volume    = {136},
  issue     = {6},
  pages     = {066703},
  numpages  = {11},
  year      = {2026},
  month     = {Feb},
  publisher = {American Physical Society},
  doi       = {10.1103/lkf9-jgv6},
  url       = {https://link.aps.org/doi/10.1103/lkf9-jgv6}
}

@article{PhysRevLett.129.247001,
  title     = {Quasiparticle Interference as a Direct Experimental Probe of Bulk Odd-Frequency Superconducting Pairing},
  author    = {Chakraborty, Debmalya and Black-Schaffer, Annica M.},
  journal   = {Phys. Rev. Lett.},
  volume    = {129},
  issue     = {24},
  pages     = {247001},
  numpages  = {6},
  year      = {2022},
  month     = {Dec},
  publisher = {American Physical Society},
  doi       = {10.1103/PhysRevLett.129.247001},
  url       = {https://link.aps.org/doi/10.1103/PhysRevLett.129.247001}
}

@article{chakraborty2024,
  title     = {Constraints on superconducting pairing in altermagnets},
  author    = {Chakraborty, Debmalya and Black-Schaffer, Annica M.},
  journal   = {Phys. Rev. B},
  volume    = {112},
  issue     = {1},
  pages     = {014516},
  numpages  = {12},
  year      = {2025},
  month     = {Jul},
  publisher = {American Physical Society},
  doi       = {10.1103/zylh-rqxl},
  url       = {https://link.aps.org/doi/10.1103/zylh-rqxl}
}

@article{PhysRevB.87.104513,
  title     = {Majorana fermions and odd-frequency {C}ooper pairs in a normal-metal nanowire proximity-coupled to a topological superconductor},
  author    = {Asano, Yasuhiro and Tanaka, Yukio},
  journal   = {Phys. Rev. B},
  volume    = {87},
  issue     = {10},
  pages     = {104513},
  numpages  = {10},
  year      = {2013},
  month     = {Mar},
  publisher = {American Physical Society},
  doi       = {10.1103/PhysRevB.87.104513},
  url       = {https://link.aps.org/doi/10.1103/PhysRevB.87.104513}
}

@article{PhysRevB.101.195303,
  title     = {From fractional solitons to {M}ajorana fermions in a paradigmatic model of topological superconductivity},
  author    = {Ziani, N. Traverso and Fleckenstein, C. and Vigliotti, L. and Trauzettel, B. and Sassetti, M.},
  journal   = {Phys. Rev. B},
  volume    = {101},
  issue     = {19},
  pages     = {195303},
  numpages  = {7},
  year      = {2020},
  month     = {May},
  publisher = {American Physical Society},
  doi       = {10.1103/PhysRevB.101.195303},
  url       = {https://link.aps.org/doi/10.1103/PhysRevB.101.195303}
}

@article{PhysRevB.101.214507,
  title     = {Bulk odd-frequency pairing in the superconducting {S}u-{S}chrieffer-{H}eeger model},
  author    = {Tamura, Shun and Nakosai, Sho and Black-Schaffer, Annica M. and Tanaka, Yukio and Cayao, Jorge},
  journal   = {Phys. Rev. B},
  volume    = {101},
  issue     = {21},
  pages     = {214507},
  numpages  = {14},
  year      = {2020},
  month     = {Jun},
  publisher = {American Physical Society},
  doi       = {10.1103/PhysRevB.101.214507},
  url       = {https://link.aps.org/doi/10.1103/PhysRevB.101.214507}
}

@article{PhysRevB.101.094506,
  title     = {Suppression of odd-frequency pairing by phase disorder in a nanowire coupled to Majorana zero modes},
  author    = {Kuzmanovski, Dushko and Black-Schaffer, Annica M. and Cayao, Jorge},
  journal   = {Phys. Rev. B},
  volume    = {101},
  issue     = {9},
  pages     = {094506},
  numpages  = {14},
  year      = {2020},
  month     = {Mar},
  publisher = {American Physical Society},
  doi       = {10.1103/PhysRevB.101.094506},
  url       = {https://link.aps.org/doi/10.1103/PhysRevB.101.094506}
}

@article{PhysRevB.100.104511,
  title     = {Signature of odd-frequency equal-spin triplet pairing in the Josephson current on the surface of {Weyl} nodal loop semimetals},
  author    = {Dutta, Paramita and Black-Schaffer, Annica M.},
  journal   = {Phys. Rev. B},
  volume    = {100},
  issue     = {10},
  pages     = {104511},
  numpages  = {10},
  year      = {2019},
  month     = {Sep},
  publisher = {American Physical Society},
  doi       = {10.1103/PhysRevB.100.104511},
  url       = {https://link.aps.org/doi/10.1103/PhysRevB.100.104511}
}

@article{PhysRevB.109.205406,
  title     = {Controllable odd-frequency {C}ooper pairs in multisuperconductor {J}osephson junctions},
  author    = {Cayao, Jorge and Burset, Pablo and Tanaka, Yukio},
  journal   = {Phys. Rev. B},
  volume    = {109},
  issue     = {20},
  pages     = {205406},
  numpages  = {10},
  year      = {2024},
  month     = {May},
  publisher = {American Physical Society},
  doi       = {10.1103/PhysRevB.109.205406},
  url       = {https://link.aps.org/doi/10.1103/PhysRevB.109.205406}
}

@article{PhysRevB.103.024501,
  title     = {Impact of impurity scattering on odd-frequency spin-triplet pairing near the edge of the {K}itaev chain},
  author    = {Mishra, Sparsh and Tamura, Shun and Kobayashi, Akito and Tanaka, Yukio},
  journal   = {Phys. Rev. B},
  volume    = {103},
  issue     = {2},
  pages     = {024501},
  numpages  = {14},
  year      = {2021},
  month     = {Jan},
  publisher = {American Physical Society},
  doi       = {10.1103/PhysRevB.103.024501},
  url       = {https://link.aps.org/doi/10.1103/PhysRevB.103.024501}
}

@article{balatsky2018oddfreq,
  title   = {Odd-frequency Pairing in Conventional Josephson Junctions},
  author  = {Alexander V. Balatsky and Sergey S. Pershoguba and Christopher Triola},
  year    = {2018},
  journal = {arXiv:1804.07244},
  url     = {https://arxiv.org/abs/1804.07244}
}
\clearpage

%------------------------------------------------------------------------
% End Matter
%------------------------------------------------------------------------
% workaround for formatting
\newlength{\savedcolumnwidth}
\setlength{\savedcolumnwidth}{\columnwidth}
\onecolumngrid
\noindent
\begin{minipage}[t]{\savedcolumnwidth}
    \section{End Matter}
    %------------------------------------------------------------------------
    % End Matter: SF junction
    %------------------------------------------------------------------------
    \textit{\label{sec:app_sf_junction}SF junction.}---For the SF junction described by Eq.~\eqref{eq:ham_sf}, we can analytically calculate the full pair amplitudes induced into the F region, given by
\end{minipage}
\begin{align}
    F_{\uparrow\uparrow}(z)                                           & =\frac{4z \Delta  t\,t_{s} \left(t_{s}^2-t^2\right) (b_z-\varepsilon_{\rm F}-\varepsilon_{\rm S})}{{\rm det}(z-H_{\rm BdG}^{\rm SF})}                                                                                                                                                                            \\
    F_{\downarrow\downarrow}(z)                                       & =\frac{4z \Delta  t\,t_{s} \left(t_{s}^2-t^2\right) (b_z+\varepsilon_{\rm F}+\varepsilon_{\rm S})}{{\rm det}(z-H_{\rm BdG}^{\rm SF})}                                                                                                                                                                            \\
    \frac12\left(F_{\uparrow\downarrow}+F_{\downarrow\uparrow}\right) & = \frac{2z \Delta b_z \left(t_s^2-t^2\right)\left(z^2-\Delta^2-t^2-t_s^2-\varepsilon_{\rm S}^2\right)}{{\rm det}(z-H_{\rm BdG}^{\rm SF})}                                                                                                                                                                        \\
    \frac12\left(F_{\uparrow\downarrow}-F_{\downarrow\uparrow}\right) & = \frac{\Delta \left(t_s^2-t^2\right)\left(\left(z^2 + b_z^2 - \varepsilon_{\rm F}^2 \right)\left(z^2-\Delta^2 - \varepsilon_{\rm S}^2 \right) - 2\left(t^2 + t_s^2 \right)\left(z^2 + \varepsilon_{\rm F} \varepsilon_{\rm S} \right) + \left(t^2 - t_s^2 \right)^2\right)}{{\rm det}(z-H_{\rm BdG}^{\rm SF})}.
\end{align}
\twocolumngrid
Here $H_{\rm BdG}^{\rm SF}$ is the full BdG Hamiltonian for the SF junction.
We can understand these results by calculating the first non-zero order contributions by using the high-frequency expansion in Eq.~\eqref{eq:f_high_freq_exp}, giving
    {\allowdisplaybreaks
        \begin{align}
            F_{\uparrow\uparrow}(z)                                           & \approx \frac{4 \Delta t\, t_s (t_s^2 - t^2)(b_z - \varepsilon_{\rm F} - \varepsilon_{\rm S})}{z^7} \\
            F_{\downarrow\downarrow}(z)                                       & \approx \frac{4 \Delta t\, t_s (t_s^2 - t^2)(b_z + \varepsilon_{\rm F} + \varepsilon_{\rm S})}{z^7} \\
            \frac12\left(F_{\uparrow\downarrow}+F_{\downarrow\uparrow}\right) & \approx \frac{2 \Delta b_z \left(t_s^2-t^2\right)}{z^5}                                             \\
            \frac12\left(F_{\uparrow\downarrow}-F_{\downarrow\uparrow}\right) & \approx \frac{\Delta \left(t_s^2-t^2\right)}{z^4}.
        \end{align}
    }
This shows that the presence of finite mixed spin-triplet pairing $(F_{\uparrow\downarrow} + F_{\downarrow\uparrow}) \neq 0$ is predicated on a magnetic field $b_z \neq 0$ as mentioned in the main text, but that equal spin pairing instead requires both spin-flip $t_s$ and spin-conserving $t$ hopping. Note that, while we here have the full expressions, due to the simplicity of the model, we could alternatively use the recursive Eqs.~(\ref{eq:rec1}-\ref{eq:rec2}) without the need to calculate the full expression. This is particularly useful for larger BdG Hamiltonians.

%------------------------------------------------------------------------
% Appendix: TMD model details
%------------------------------------------------------------------------
\textit{\label{sec:tmdmodel} Proximitized TMD monolayer.}---
\begin{figure}[!h]
    \centering
    \includegraphics[width=0.6\columnwidth]{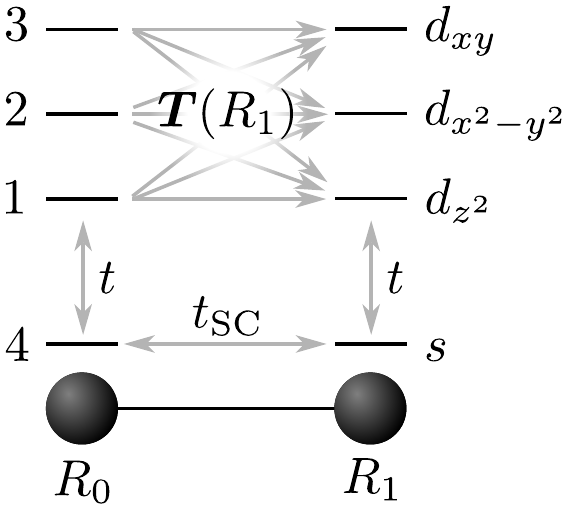}
    \caption{Sketch illustrating the nearest-neighbor hopping amplitudes $\bm{T}(R_1)$ between the central site $R_0$ and the nearest-neighbor site $R_1$, with the labeling of the TMD and SC orbitals with indices $i=1,2,3,4$. }
    \label{fig:orbital_levels_2atoms}
\end{figure}
%To illustrate the usefulness of our derived conditions on odd-frequency pairing, we apply them to proximity-induced superconductivity in a monolayer of a transition metal dichalcogenide (TMD), as discussed in Sec.~\ref{sec:application1} of the main text.
For a wide range of TMDs, proximity-induced superconductivity can be modeled using the Hamiltonian $\hat{h}_{\bm{k}}$ in Eq.~\eqref{eq:hamk_example1}, where the first three-orbital block reproduces the low-energy band structure of the TMD monolayer following \cite{liu_three-band_2013}, based on the $d_{z^2}$, $d_{xy}$, and $d_{x^2-y^2}$ orbitals of the transition metal atoms.
The superconductor is modeled by an additional $s$-wave orbital with dispersion $H_{\bm{k}}^{44}$.
The nearest-neighbor hopping elements of the underlying tight-binding model are illustrated in Fig.~\ref{fig:orbital_levels_2atoms}, between the central atom $R_0$ and one of its nearest neighbors $R_1$.
The explicit form of the matrix elements in Eq.~\eqref{eq:hamk_example1} are given as \cite{liu_three-band_2013, kayatz2026tmd}
{\allowdisplaybreaks
    \begin{align*}
        H^{11}_{\bm{k}} & = \varepsilon_1 + 2t_0 (\cos2\alpha + 2 \cos\alpha \cos\beta)                     \\
                        & + 2r_0 (\cos2\beta + 2 \cos3\alpha \cos\beta)                                     \\
                        & + 2u_0 (\cos4\alpha + 2 \cos2\alpha\cos2\beta)                                    \\
        H^{12}_{\bm{k}} & = 2it_1(\sin2\alpha + \sin\alpha\cos\beta) - 2\sqrt{3}t_2\sin\alpha\sin\beta      \\
                        & + 2 \sin3\alpha ((r_1+r_2) \sin\beta + i (r_1 - r_2)\cos\beta)                    \\
                        & + 2iu_1(\sin4\alpha + \sin2\alpha\cos2\beta) - 2\sqrt{3}u_2\sin2\alpha\sin2\beta  \\
        H^{13}_{\bm{k}} & = 2t_2(\cos2\alpha - \cos\alpha\cos\beta) + 2i \sqrt{3} t_1 \cos\alpha\sin\beta   \\
                        & - \frac{2}{\sqrt{3}} (r_1 + r_2) (\cos3\alpha \cos\beta - \cos2\beta)             \\
                        & + \frac{2}{\sqrt{3}} i(r_1 - r_2) \sin\beta (\cos3\alpha + 2\cos\beta)            \\
                        & + 2u_2(\cos4\alpha - \cos2\alpha\cos2\beta) + 2i\sqrt{3}u_1 \cos2\alpha\sin2\beta \\
        H^{14}_{\bm{k}} & = t                                                                               \\
        H^{22}_{\bm{k}} & = \varepsilon_2 + 2 t_{11} \cos2\alpha + (t_{11} +3t_{22}) \cos\alpha\cos\beta    \\
                        & + 4r_{11} \cos3\alpha\cos\beta + 2(r_{11} + \sqrt{3} r_{12})\cos2\beta            \\
                        & + 2u_{11} \cos4\alpha + (u_{11} +3u_{22}) \cos2\alpha\cos2\beta                   \\
        H^{23}_{\bm{k}} & = 4i t_{12} \sin\alpha(\cos\alpha - \cos\beta)                                    \\
                        & + \sqrt{3} (t_{22} - t_{11}) \sin\alpha\sin\beta                                  \\
                        & + 4r_{12}\sin3\alpha\sin\beta                                                     \\
                        & + 4iu_{12} \sin2\alpha (\cos2\alpha - \cos2\beta)                                 \\
                        & + \sqrt{3}(u_{22} - u_{11})\sin2\alpha\sin2\beta                                  \\
        H^{33}_{\bm{k}} & = \varepsilon_2 + 2 t_{22} \cos2\alpha + (3 t_{11} + t_{22}) \cos\alpha \cos\beta \\
                        & + 2r_{11}(\cos2\beta + 2 \cos3\alpha\cos\beta)                                    \\
                        & + \frac{2}{\sqrt{3}} r_{12} (4 \cos3\alpha \cos\beta - \cos2\beta)                \\
                        & + 2u_{22}\cos4\alpha + (3u_{11}+u_{22})\cos2\alpha\cos2\beta                      \\
        H^{44}_{\bm{k}} & = 2 t_{\text{SC}} (\cos2\alpha + 2\cos\alpha\cos\beta) - \mu_{\text{SC}},
    \end{align*}
}
where $t_i$, $r_i$, and $u_i$ denote the nearest-, second-nearest-, and third-nearest-neighbor hopping amplitudes, respectively, following the notation of \cite{liu_three-band_2013}.
The on-site energies are given by $\varepsilon_i$, $t$ is the interlayer coupling and $t_{\text{SC}}$ and $\mu_{\text{SC}}$ are the hopping strength and chemical potential of the superconductor.
The momentum dependence is set by $\alpha = k_x a / 2$, $\beta = \sqrt{3} k_y a / 2$ with lattice constant $a$.
The remaining matrix elements can be obtained from hermiticity, $H^{ij}_{\bm{k}} = (H^{ji}_{\bm{k}})^*$.
For a more detailed discussion of the superconducting properties of this system, as well as realistic values for the hopping strengths of specific TMDs such as $\text{MoS}_2$, we refer to \cite{liu_three-band_2013, kayatz2026tmd}.

\end{document}